\documentstyle[epsf,multicol,aps]{revtex}
\begin{document}
\draft
\title{Near-Critical Gravitational Collapse and the Initial Mass
Function of Primordial Black Holes}
\author{J.C. Niemeyer}
\address{University of Chicago, Department of Astronomy and
Astrophysics, 5640 South Ellis Avenue, Chicago, IL 60637} 
\author{K. Jedamzik}
\address{Max-Planck-Institut f\"ur Astrophysik,
Karl-Schwarzschild-Str. 1, 85740 Garching, Germany} 
\maketitle

\begin{abstract}
The recent discovery of critical phenomena arising in gravitational
collapse near the threshold of black hole formation is used to
estimate the initial mass function of primordial black holes (PBHs). It is
argued that the scaling relation between black hole mass and
initial perturbation found for a collapsing radiation fluid in an
asymptotically flat space-time also applies to PBH formation in a
Friedmann universe, indicating the  possible formation
of PBHs with masses much smaller than one horizon mass. Owing to
the natural fine-tuning of initial conditions by the exponential decline
of the probability distribution for primordial density fluctuations,
sub-horizon mass PBHs are expected to form at all
epochs. This result suggests that the
constraints on the primordial fluctuation spectrum based on the
abundance of PBHs at different mass scales may have to be revisited.
\end{abstract}


\begin{multicols}{2}
In this Letter, we consider the initial mass function (IMF) of
primordial black holes (PBHs) formed in the process of gravitationally
collapsing primordial density fluctuations in the radiation dominated
phase of the early universe \cite{PBHorigin,carhaw74}. Implications of the PBH
number and mass spectrum with regard 
to their contribution to the cosmic density and the $\gamma$-ray
background (via Hawking evaporation) have been employed to constrain the
spectral index of the primordial fluctuation spectrum
\cite{carrea94,green97}. Two aspects play a central role in these
studies: first, for each horizon-sized space-time region there exists
a critical threshold value, $\delta_{\rm c}$, for the density (or mass)
contrast $\delta$, separating its further evolution between formation of
a black hole ($\delta > \delta_{\rm c}$) and dispersion by pressure
forces ($\delta < \delta_{\rm c}$) (we shall use the
term ``horizon'' to denote the particle horizon, $r_{\rm h} \sim
t$). Comparing the Jeans and horizon lengths at the time when the
collapsing region breaks away from Hubble expansion, one finds that
$\delta_{\rm c}$ must be of order unity \cite{carr75}. The second key
assumption relates to the final mass of the black hole, $M_{\rm bh}$.  
It is commonly assumed that $M_{\rm bh}$ is approximately
equal to the mass of the collapsing region and thus to the horizon mass
at the epoch of formation,  $M_{\rm h}$.
Nevertheless, detailed predictions for the PBH IMF have not
previously been made. Based on a scaling relation discovered in
gravitational collapse of various near-critical space-times,
generalized to collapsing density
perturbations in an Einstein-de Sitter universe, we are able
to derive a universal, two-parameter PBH IMF, 
applicable when PBH number densities are dominated from
fluctuations collapsing during one particular epoch. 
Here the two parameters
in the PBH IMF carry all the information of the statistics
of the initial density spectra and the perturbation shapes.
We show that when the perturbation overdensity
is sufficiently close to the critical overdensity for PBH formation,
$\delta_{\rm c}$, the final mass of the resulting PBH may be an
arbitrarily small fraction of the horizon mass, providing a conceptual
difference to our current understanding of the process of PBH formation.  

It is possible that PBH formation is the only natural example
for critical phenomena in gravitational collapse, a field of
considerable interest in classical general relativity that was previously
believed to have no astrophysical application. 
Triggered by the intriguing results of Choptuik \cite{chop93} who demonstrated
scaling and self-similarity in the gravitational collapse of a massless
scalar field near the threshold of black hole formation, critical
phenomena were studied for a number of different set-ups, including
spherically symmetric radiation fluids \cite{evans94}, Yang-Mills
fields \cite{chopea96}, and axially symmetric collapsing gravity waves
\cite{abr93}. In all cases, families of initial data quantified by a
single generic parameter $\delta$ were found to give rise to a scaling
relation of the form
\begin{equation}
\label{critcoll}
M_{\rm bh}(\delta) = K \left( \delta - \delta_{\rm c} \right)^\gamma
\end{equation}
near the critical point for black hole formation, $\delta_{\rm
c}$. The specific choice of $\delta$ is arbitrary since differentiable
transformations of
$\delta$ leave (\ref{critcoll}) invariant, changing only the constant $K$
to leading order \cite{gund96}. Another noteworthy feature of
near-critical solutions is the the appearance of discrete (scalar
field collapse) or continuous (perfect fluid collapse)
self-similarities.  

Equation (\ref{critcoll}) is, in general, irrelevant for the formation of
astrophysical black holes. Degeneracy pressure of neutrons or
electrons introduces intrinsic limiting mass scales 
of hydrostatic stability, such as the
Chandrasekhar mass, violating the scale-free behavior 
indicated by equation (\ref{critcoll}). Moreover, equation
(\ref{critcoll}) is only valid in the immediate neighborhood of
$\delta_{\rm c}$, requiring a high degree of fine-tuning of the
initial conditions which is unnatural under most circumstances.
In PBH formation, on the other hand, it is expected that
most regions collapsing to a black hole will have overdensities
close to the critical overdensity for PBH formation, $\delta_{\rm c}$,
owing to a steeply declining probability distribution for primordial
density fluctuations. Typical cosmic initial conditions thus provide
the fine-tuning of initial conditions, required for near-critical
collapse. Further, with the exception of cosmological phase
transitions that will not be considered here, the matter
collapsing to PBHs is well described by a perfect fluid with radiation
dominated equation of state,  $p=\rho /3$, where $p$ and $\rho$
are pressure and energy density, respectively. Hence, the problem for
PBH formation in radiation dominated 
cosmological epochs and the perfect fluid collapse studied numerically
by Evans and Coleman (EC) \cite{evans94} differ only with regard to the
background space-time. While canonical initial conditions
for PBH formation involve curvature perturbations in an expanding
Friedmann-Robertson-Walker (FRW) space-time, EC used initial data
embedded in in an asymptotically stationary space-time for their
collapse simulations. 

In addition to their numerical simulations, EC found a self-similar
solution to  
the equations of motion and gravitation in the limit $\delta \to
\delta_{\rm c}$. A self-similar ansatz reduces the spatial and
temporal degrees of freedom to a single self-similar coordinate and
thereby transforms the system of partial differential equations into
ordinary ones. Demanding regularity at the center and along the
ingoing acoustic characteristic, corresponding to the absence of a
shock, the system of ODEs can be solved and the solution coincides
well with their numerical results. As suggested by EC, the critical
exponent of (\ref{critcoll}) was subsequently derived by analyzing linear
perturbations of the self-similar solution: Koike, Hara and Adachi
\cite{koike95} obtained 
$\gamma = 0.3558019$ for a collapsing radiation fluid. Note that
neither the self-similar solution nor the perturbation analysis rely
on asymptotic flatness of the space-time; on the contrary, EC's
self-similar solution is not asymptotically flat. As EC's solution
converges neither to a flat stationary space-time nor to an exact FRW
solution, it invariably breaks down at large radii for both
asymptotic behaviors.

The main reason to expect the emergence of self-similarity in near-critical
gravitational collapse occuring in asymptotically FRW space-times is
the separation of characteristic scales: Just
as in the asymptotically stationary case studied by EC, the solution
forms an intermediate asymptotic between two widely separated length
scales \cite{bar72}. The scale $r_0$ of the fluid perturbation
$\delta$ at the onset of collapse is given by $\delta^{-1/2} r_{\rm
h}$ \cite{carr75} if the initial perturbation amplitude is evaluated at horizon
crossing. $r_0$ can be identified with the transition
from Hubble expansion of the asymptotic FRW space-time to the
collapse-dominated region $r < r_0$. On small scales, deviation from exact
criticality leads to violation of self-similarity if $r$ approaches $r_1 \sim 
K |\delta - \delta_{\rm c}|$ \cite{com3}. The ratio $r_0/r_1$ can be made
arbitrarily large by chosing initial data close to the critical
point. In the limit $\delta \to \delta_{\rm c}$, we therefore assume
that gravitational collapse of a radiation fluid is well described by
the self-similar solution of EC \cite{evans94} and the critical
exponent $\gamma \approx 0.356$ \cite{koike95}, independent of the asymptotic
behavior of the background space-time. 
We note that preliminary results of numerical simulations of the PBH formation
process in the early universe confirm (1) the scaling relation 
and (2) the applicabilty of scaling for commonly assumed
parameters of the statistics of pre-existing cosmic
density fluctuations (see below). 
The results of this numerical investigation
will be presented elsewhere \cite{niejed98}. 

Based on the arguments above we will henceforth employ equation 
(\ref{critcoll}) for the masses of
PBHs formed by collapsing primordial density perturbations slightly
exceeding $\delta_{\rm c}$, with an exponent $\gamma \approx 0.356$
independent of initial perturbation shape.
We assume a Gaussian probability
distribution for density fluctuations entering the horizon, 
\begin{equation}
\label{pdelta}
P(\delta) = \frac{1}{\sqrt{2\pi} \sigma} \, \exp{\left(-\frac{\delta^2}{2
\sigma^2}\right)}\,\,,
\end{equation}
where $\sigma$ is the, possibly scale-dependent, root-mean-square 
fluctuation amplitude. Equation (\ref{pdelta}) allows us to compute the
fraction of horizon-sized regions collapsing to PBHs at a given
epoch \cite{carr75}
\begin{equation}
\beta = \int_{\delta_{\rm c}}^1 P(\delta) \, d\delta \approx {\sigma}
\exp{\left(-\frac{\delta_{\rm c}^2}{2 \sigma^2}\right)}\,\,.
\end{equation}
The upper integration limit reflects that if $\delta > 1$, the
collapsing space-time region corresponds to a separate
closed universe instead of a black hole \cite{carhaw74}, and
the approximation on the right hand side is valid to within a factor
of a few for $\sigma /\delta_{\rm c} \le 0.2$. 
It is noted that non-Gaussian effects may be important for
$\delta_{\rm c} \gg \sigma$ \cite{bull96}, but a Gaussian distribution
suffices for the demonstration purpose of this work.

In what follows we assume that cosmological PBH formation is dominated
by perturbations of one particular length scale, defining a
characteristic epoch of PBH formation by the time the perturbations
cross into the horizon. 
Such an analysis should be adequate when either the initial
perturbation spectrum exhibits a peak on a given scale \cite{yoko95},
or PBH formation is most probable during a specific epoch by virtue of
the equation
of state \cite{jeda97}. It is also approximately valid for blue
initial perturbation spectra where PBH formation is most efficient on the
smallest scale under consideration \cite{com2}. 

With these assumptions, and approximating
incorrectly for the moment that the mass of the resulting PBH is $M_{\rm h}$,
we may compute the value of the PBH mass density divided by the cosmic
background density, 
\begin{eqnarray}
\label{omold}
\hat \Omega_{\rm pbh,old} \equiv \langle\frac{\rho_{\rm
bh}}{\rho_0}\rangle & = & M_{\rm h}^{-1}\int_{\delta_{\rm c}}^1 M_{\rm 
bh}\,P(\delta) \, d\delta \nonumber \\
&\approx &\beta \quad \mbox{for} \quad M_{\rm bh} \approx M_{\rm h}\,\,,
\end{eqnarray}
where the hat indicates that $\hat \Omega_{\rm pbh}$ is evaluated at the
time of PBH formation. 

As a straightforward modification of $\hat \Omega_{\rm pbh}$ , we
use the continous distribution of PBH masses (\ref{critcoll}) in
(\ref{omold}) and 
re-evaluate the integral. Doing so, we implicitly assume that
(\ref{critcoll}) is valid for $\delta$ as large as unity; this need not
necessarily be the case. However, the largest contribution to the
integral comes from $\delta \approx \delta_{\rm c}$ owing to the 
exponential form
of $P(\delta)$, and thus our assumption is justified. The integrand
rises steeply to a maximum at  
\begin{equation} 
\delta_{\rm m} = \frac{1}{2}\left(\delta_{\rm c} + \sqrt{4 \gamma
\sigma^2 + \delta_{\rm c}^2}\right) = \delta_{\rm c} +
\frac{\gamma \sigma^2}{\delta_{\rm c}} +
{\rm O}(\sigma^4)
\end{equation}
close to the lower integration boundary, and the black hole mass at
this point is 
\begin{equation}
\label{mbh}
M_{\rm bh}(\delta_{\rm m}) = K\left(\frac{\gamma
\sigma^2}{\delta_{\rm c}}\right)^\gamma \approx k\sigma^{2\gamma}\,
M_{\rm h}\,\, ,
\end{equation}
with the dimensionless $k$ defined by $K=k M_{\rm h}$.
The modified expression for $\hat \Omega_{\rm pbh}$ is thus
\begin{eqnarray}
\label{omnew}
\hat \Omega_{\rm pbh,new} &= &  M_{\rm h}^{-1}\int_{\delta_{\rm c}}^1 M_{\rm
bh}(\delta)\,P(\delta) \, d\delta  \nonumber \\
&\approx &  M_{\rm h}^{-1}\int_{\delta_{\rm m}}^1 M_{\rm
bh}(\delta)\,P(\delta) \, d\delta  \nonumber \\
&\approx &
k\sigma^{1+2\gamma}\exp{\left(-\frac{\delta_{\rm c}^2}{2 \sigma^2}\right)}
\approx k\sigma^{2 \gamma} \beta\,\,,
\end{eqnarray}
where the integral was asymptotically expanded to first order. 
Equation (\ref{omnew}) shows that the average black hole
mass produced at each epoch is approximately given by (\ref{mbh}),
since
\begin{equation}
\label{mbhmean}
\langle M_{\rm bh} \rangle = \beta^{-1} \int_{\delta_{\rm c}}^1 M_{\rm
bh}(\delta)\,P(\delta) \, d\delta \approx  k\sigma^{2 \gamma} M_{\rm
h}\,\,. 
\end{equation}

We can now determine the PBH initial mass function (IMF) when
PBH number densities are dominated from formation during one particular
epoch. The global PBH mass spectrum generally involves an
integration over all epochs, a formidable problem,
which will not be attempted here. We define
the PBH IMF as the fraction $d\phi$ of PBH number per logarithmic mass
interval, normalized such that
\begin{equation}
\label{norm}  
\int_{-\infty}^{{\rm ln}M_{\rm bh}(\delta =1)}
{d\phi\over d({\rm ln}M^{\prime}_{\rm bh})}
d({\rm ln}M^{\prime}_{\rm bh}) = 1\,\, .
\end{equation}
This mass function is given by
\begin{eqnarray}
\label{phi}
{d\phi\over d({\rm ln}M_{\rm bh})}& = & \beta^{-1}P(\delta (M_{\rm bh}))
{d\delta\over d({\rm ln}M_{\rm bh})} \nonumber \\
&= &{1\over\sqrt{2\pi} \beta\sigma\gamma} m_{\rm bh}^{{1\over\gamma}}{\rm
exp}\biggl(-{\bigl(\delta_{\rm c}+m_{\rm bh}^{1\over\gamma}\bigr)^2
\over 2\sigma^2}\biggr) \,\, ,
\end{eqnarray}
where $m_{\rm bh}$ is black hole mass
in units of $k M_{\rm h}$, and where we have used equation (\ref{pdelta})
for $P(\delta )$. 
The PBH IMF of equation (\ref{phi}) has wider applicability
than naively thought. Imagine PBH formation in the case of
non-Gaussian statistics, in particular, when $P(\delta )$ is
different from equation (\ref{pdelta}). In this case one may
search for a control parameter $\delta^{\prime}(\delta )$ which renders
$P(\delta )d\delta /d\delta^{\prime}$ Gaussian. 
Applying this transformation between control parameters 
to equation (\ref{critcoll}), one will obtain a form-invariant equation
(\ref{critcoll}) with modified constants $K^{\prime}$ and $\delta_{\rm c}^{\prime}$,
provided the limit of near-critical gravitational collapse still applies. 
Equation (\ref{phi}) then defines a universal two-parameter
family of PBH IMFs, applicable for many initial conditions, with
the parameters $K$ and $\delta_{\rm c}$ carrying all the information
about the statistics of initial conditions and the shapes of
perturbations.

The mass function of equation (\ref{phi}) exhibits a maximum at
\begin{equation}
\label{Mmax}
M_{\rm bh}^{\rm max} = k \left(\frac{\sigma^2}{\delta_{\rm
c}}\right)^{\gamma} M_{\rm h} \,\, ,
\end{equation}
which approximately equals the average black hole mass
equation (\ref{mbhmean}). PBHs in
cosmological interesting numbers are formed during the
evolution of the early universe for values of $\sigma /\delta_{\rm c}\approx
0.1 - 0.2$,  provided Gaussian statistics holds
\cite{green97}. 
Such values of $\sigma /\delta_{\rm c}$ yield typical volume
collapse fractions in the range $\beta\approx 10^{-6} - 10^{-23}$ and,
depending on the epoch of formation, imply PBH number
densities significantly contributing
to the present mass density, or the $\gamma$-background. Inserting $k
\approx 3.3$ found by EC \cite{com1}, $\sigma /\delta_{\rm c} \approx 0.15$, and
$\delta_{\rm c} \approx 1/3$ \cite{carr75}, we find 
\begin{equation}
M_{\rm bh}^{\rm max} \approx 0.6 M_{\rm h}\,\,.
\end{equation} 

It is not surprising that the maximum of the IMF at a fixed epoch
coincides with the horizon mass to within an order of magnitude, since
the latter
determines the mass scale for collapse. However, depending on the
value of $\sigma$, a fraction of all PBHs formed at each epoch will
have masses significantly smaller than $M_{\rm h}$, implying a
fundamental conceptual difference between 
this work and previous calculations. It was previously assumed that
there exists a
one-to-one correspondence between $M_{\rm bh}$ and redshift
$z$. Under this assumption, it was straightforward to relate $M_{\rm bh}$ to
a single energy scale, i.e. microscopically small black holes only
formed at very early times. Using equation (\ref{critcoll}) instead, this
simplification is no longer valid; the formation of
black holes with a continuous IMF allows the
formation of microscopic PBHs at all epochs. 
The formulation of observational constraints based on these 
results, such as constraints on the spectral index
of initial density spectra, requires a detailed analysis of 
the PBH IMF integrated over all epochs which is beyond the scope of
this Letter.

The authors wish to thank S.E. Woosley for helpful discussions.

\end{multicols}


\begin{thebibliography}{99}
\bibitem{PBHorigin} Ya. B. Zeldovich, and I. D. Novikov,
Sov. Astron. A. J. {\bf 10}, 602 (1967); S. W. Hawking, M.N.R.A.S. {\bf
152}, 75 (1971).

\bibitem{carhaw74} B. J. Carr and S. W. Hawking, M.N.R.A.S. {\bf 168},
399 (1974).

\bibitem{carrea94} B. J. Carr, J. H. Gilbert, and J. E. Lidsey,
Phys. Rev. D. {\bf 50}, 4853 (1994)

\bibitem{green97} A. M. Green and A. R. Liddle, astro-ph/9704251.

\bibitem{carr75}  B. J. Carr, Astrophys. J. {\bf 201}, 1 (1975).

\bibitem{chop93} M. W. Choptuik, Phys. Rev. Lett. {\bf 70}, 9 (1993).

\bibitem{evans94} C. R. Evans and J. S. Coleman, Phys. Rev. Lett {\bf
72}, 1782 (1994). 

\bibitem{chopea96} M. W. Choptuik, T. Chmaj, and P. Bizon,
gr-qc/9603051 (1996).

\bibitem{abr93} A. M. Abrahams and C. R. Evans, Phys. Rev. Lett. {\bf
70}, 2980 (1993).

\bibitem{gund96} C. Gundlach, gr-qc/960623 (1996).

\bibitem{koike95} T. Koike, T. Hara, and S. Adachi,
Phys. Rev. Lett. {\bf 74}, 5170 (1995).

\bibitem{bar72} G. Barenblatt and Ya. B. Zel'dovich, Ann Rev. Fluid
Mech. {\bf 4}, 285 (1972). 

\bibitem{com3} We use $G=c=1$ throughout the paper such
that K has dimensions of mass and length. For example, the
horizon mass is: $M_{\rm h}=(4\pi /3)\rho r_{\rm h}^3=r_{\rm h}/2$.

\bibitem{niejed98} J. C. Niemeyer and K. Jedamzik, in preparation
(1998).

\bibitem{bull96} J. S. Bullock and J. R. Primack, astro-ph/9611106
(1996).

\bibitem{yoko95} J. Yokoyama, Astronomy \& Astrophysics {\bf 318}, 673
(1997).

\bibitem{jeda97} K. Jedamzik, Phys. Rev. D {\bf 55}, R5871 (1997).

\bibitem{com2}
Clearly, this analysis is incorrect
for scale-invariant perturbation spectra. Nevertheless, given
the COBE normalization of $\sigma\approx 10^{-5}$ on large scales and
the assumption of Gaussianity, scale-invariance would imply that
PBH formation in the early universe is of no cosmological importance.

\bibitem{com1}
Doing so, we make a specific choice for the control parameter
$\delta$. Evans and Coleman defined $\delta = 2 M/ r_0$. 
This corresponds in our case to the mass that is 
contained within the horizon sphere
in addition to the unperturbed FRW horizon mass. We note here that for
canonical Gaussian 
density perturbations $\delta$ in equation (\ref{pdelta}) is the
average overdensity in uniform Hubble constant gauge determined
in the limit of linear evolution, which may yield a different value of
$k$. Further, $k$ may be perturbation shape dependent such that
an exact evaluation of this quantity has to await the
results of numerical simulations.


\end{thebibliography}
\end{document}